# Computationally Discovered Potentiating Role of Glycans on NMDA Receptors

Anton V. Sinitskiy,[1] Nathaniel H. Stanley,[2,3] David H. Hackos,[4] Jesse E. Hanson,[4] Benjamin D. Sellers,[3] & Vijay S. Pande[1,5]

[1] Department of Chemistry, Stanford University, Stanford, California 94305
[2] Stanford ChEM-H, Stanford University, Stanford, California 94305
[3] Department of Discovery Chemistry, Genentech, Inc., 1 DNA Way, South San Francisco, CA 94080, USA
[4] Department of Neuroscience, Genentech, Inc., 1 DNA Way, South San Francisco, CA 94080, USA
[5] Department of Computer Science and Department of Structural Biology, Stanford University, Stanford, California 94305

Correspondence and requests for materials should be addressed to A.V.S. (email: sinitskiy@stanford.edu) and V.S.P. (email: pande@stanford.edu).

N-methyl-D-aspartate receptors (NMDARs) are glycoproteins in the brain central to learning and memory. The effects of glycosylation on the structure and dynamics of NMDARs are largely unknown. In this work, we use extensive molecular dynamics simulations of GluN1 and GluN2B ligand binding domains (LBDs) of NMDARs to investigate these effects. Our simulations predict that intra-domain interactions involving the glycan attached to residue GluN1-N440 stabilize closed-clamshell conformations of the GluN1 LBD. The glycan on GluN2B-N688 shows a similar, though weaker, effect. Based on these results, and assuming the transferability of the results of LBD simulations to the full receptor, we predict that glycans at GluN1-N440 might play a potentiator role in NMDARs. To validate this prediction, we perform electrophysiological analysis of full-length NMDARs with a glycosylation-preventing GluN1-N440Q mutation, and demonstrate an increase in the glycine EC50 value. Overall, our results suggest an intramolecular potentiating role of glycans on NMDA receptors.

**INTRODUCTION**

N-methyl-D-aspartate receptors (NMDARs) are transmembrane ion channels expressed in the nervous system and other organs. Malfunction of NMDARs is implicated in the pathology of various disorders, including schizophrenia, epilepsy, intellectual disability and autism.[1-4] Each NMDAR consists of four subunits: two GluN1 subunits, and two GluN2 or GluN3 subunits. A large number of variants of NMDARs exists *in vivo*, arising from combinations of subunit types (GluN2A-D, GluN3A-B) and splicing variants (eight variants for GluN1, two variants for GluN3A).[1,2]

NMDARs, like most membrane proteins, are heavily glycosylated. At least 11 glycans are attached to GluN1, at least 4 glycans to GluN2A, and at least 7 glycans to the GluN2B subunit.[5-7] Most of the glycans in NMDARs seem to be high-mannose $Man_5GlcNAc_2$ (Man5) glycans,[8] though other type of glycans may also occur.[5-7] Data on the amount of glycosylation of an NMDAR are partially contradictory, but imply that the sites are nearly a hundred percent occupied by glycans.[7-12] Surprisingly, glycans attached to NMDARs have not received much attention in previous publications on the structure and function of NMDARs.

The effect of site-specific glycosylation on the structure and dynamics of NMDARs has not been investigated, and the neurological and psychiatric implications of abnormal NMDAR glycosylation patterns are unknown.[13] The removal of *all* glycans from NMDARs was reported to decrease EC50 for glutamate by a factor of 3.6±0.7,[10] increase the dissociation constant for non-competitive antagonist MK801 by a factor of 4.4±1.4,[9] and reduce the ratio of the steady-state current amplitudes induced by 50 μM and 1 mM NMDA by a factor of 1.3±0.1.[13] Treatment of NMDARs with certain lectins (glycan-binding proteins) increases EC50 for NMDA by 61-88%.[7] Consequences of changes in the glycosylation state at *specific* sites on NMDAR properties, however, remain poorly investigated.[13] While no correlation between the overall level of NMDAR glycosylation and schizophrenia has been found,[12] one hundred glycosylation disorders are known, including disorders with neurological symptoms, such as psychomotor retardation, ataxia, and hypotonia.[14]

NMDARs consist of relatively autonomous functional parts or domains, as demonstrated by electrophysiological and pharmacological studies of chimeric NMDARs.[15,16] The modular character of NMDARs has been widely used in the previous work on NMDARs, for example, in the reconstruction of atomistic structures of NMDARs in various functional states from cryoEM data[17,18] and in computational studies of NMDARs.[19-22] In this paper, we follow this approach and focus on the ligand-binding domains (LBDs) of the GluN1 and GluN2B subunits of NMDARs. These modules, 292 and 295 amino acid residues in size respectively, collectively comprise nearly one fourth of the full receptor (GluN1/GluN2B isoform) (Fig. 1). Each NMDAR includes two copies of each of these domains. Coagonists glycine or D-serine bind to GluN1 LBD, and the agonist glutamate binds to GluN2B LBD. Binding (or unbinding) of agonists or coagonists is believed to result in a conformational change in the corresponding domain, namely clamshell-like closing (or opening) of the domain (Fig. 2).[20,23-27] If three events occur simultaneously: (1) glycine or D-serine binds to GluN1 LBD, (2) glutamate binds to GluN2 LBD, and (3) the magnesium 'plug' is released from the transmembrane domain (TMD) by an appropriately depolarized membrane voltage, then the ion channel pore opens and calcium cations enter the cell, triggering signal cascades responsible for synaptic plasticity.[1] Disruptions in D-serine and glycine binding to GluN1 LBD have implications in schizophrenia.[28,29] Our investigation of GluN1 and GluN2B LBDs of NMDAR could clarify the connection between the (de)glycosylation of the full NMDARs and their biomedically relevant properties.

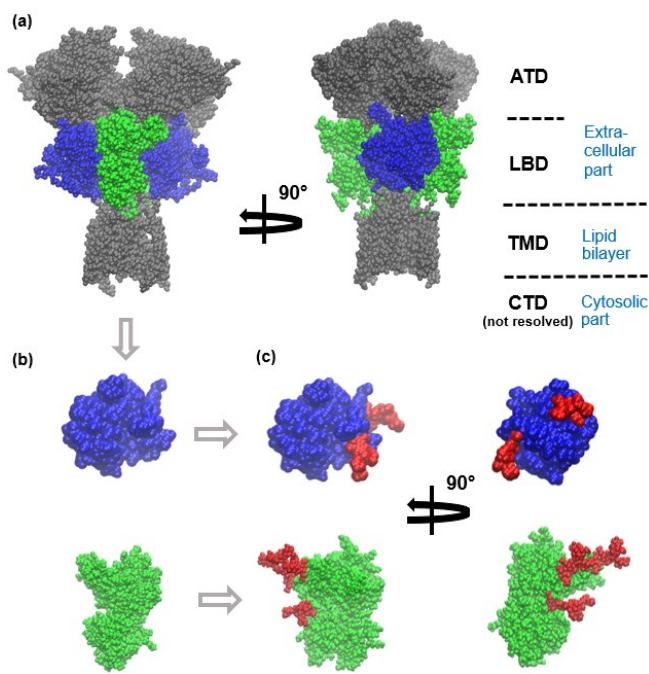

**Fig. 1.** (a) Ligand binding domains (LBD) of GluN1 (*blue*) and GluN2B (*green*) subunits are parts of an NMDA receptor (*gray, blue* and *green*; protein part only shown). Each NMDAR contains two GluN1 and two GluN2B LBDs. The overall architecture of NMDARs is annotated on the right: the amino terminal domain (ATD) and the LBD are extracellular parts, the transmembrane domain (TMD) is immersed into the lipid bilayer, and the carboxyl terminal domain (CTD, not resolved in X-ray structures) is a cytosolic part. (b) For computational feasibility, this work focuses on the independent GluN1 LBD and GluN2B LBD. (c) Three Man5 glycans (*red*) were attached to GluN1 LBD and three Man5 glycans to GluN2B LBD to match the glycosylation pattern of NMDARs in the brain.

In this paper, we adopt a novel approach to studying the consequences of glycosylation of NMDARs, namely computer simulations at atomic resolution, followed by experimental verification. In the past, computational modeling has played an indispensable role in the understanding of folding and conformational transitions in polypeptides and small proteins.[30] Simulating proteins with ~200-300 amino acid residues on biologically relevant timescales (up to ms) has recently become possible due to increases in computational power.[31,32]. The present work differs from previous simulations of NMDARs or their parts[19-21,33-37] in that the simulated systems include glycans, and the aggregate duration of molecular dynamics (MD) trajectories (0.6 milliseconds) exceeds that in the previous works by at least two orders of magnitude, closing the gap between the physiologically relevant and simulated timescales Quantitative statistical analysis based on Markov state models (MSMs) allows us to deduce equilibrium properties of the modeled systems from finite-length MD trajectories. Finally, our key prediction following from the simulations, namely the potentiator role of specific glycans on NMDARs, is corroborated by voltage-clamp electrophysiology experiments on wild-type and mutant full-length NMDARs.

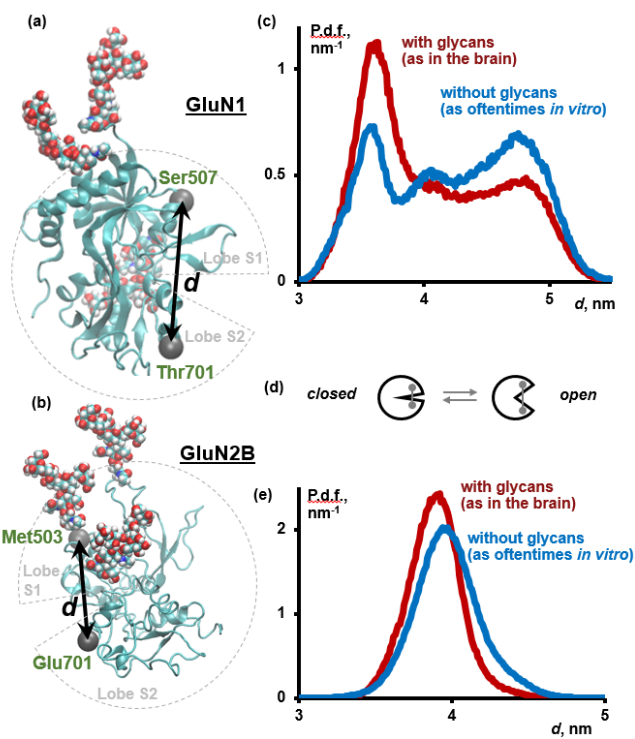

**Fig. 2.** (a-b) The physiologically most important conformational changes in the GluN1 (a) and GluN2B (b) LBDs are believed to be clamshell-like opening/closing motions, which can be quantified, for example, by changes in the distance $d$ between $C_\alpha$ atoms in residues 507 and 701 in GluN1 or residues 503 and 701 in GluN2B (*gray, van-der-Waals spheres*). Protein is shown in *cartoon representation*; glycans, *van-der-Waals spheres*. (c) Glycosylation stabilizes closed conformations of the GluN1 LBD, though open conformations are still populated. (P.d.f.: probability distribution function.) (d) A cartoon representation of the clamshell-like opening/closing motion in LBDs, with open conformations corresponding to larger values of $d$ in panels (c,e). (e) Glycosylation of the GluN2B LBD stabilizes closed-clamshell conformations as well, though this effect is less pronounced as in GluN1 LBD.

## RESULTS

**Glycosylation stabilizes closed-clamshell conformations of GluN1 LBD and GluN2B LBD.** Our simulations predict that both glycosylated and non-glycosylated GluN1 LBDs populate a wide spectrum of conformations at equilibrium, ranging from far open to far closed ones (Fig. 2). This result suggests that the available X-ray structures of GluN1 LBD may not be capturing the full variety of conformations possible for the GluN1 LBD. Recently, cryoEM studies of NMDARs revealed several distinct conformations corresponding to the same functional state (agonist-bound non-active[17] and agonist-and-antagonist-bound[18]). Our results demonstrate, however, that a small number of discrete conformations (as four and six in the two cited papers, respectively) may be insufficient for a faithful representation of the conformational heterogeneity of NMDARs under physiological conditions.

As for the changes incurred by glycosylation, we have found that the glycosylated GluN1 LBD tends to more frequently visit closed states at equilibrium, though open states are also accessible. Non-glycosylated GluN1 LBD does not demonstrate this preference and populates open and closed conformations to a near-equal extent (Fig. 2c). Bootstrapping[38] demonstrates that the conformations with the interlobe distance $d$ in the range of 3.6 to 3.8 nm are statistically significantly more populated, and those with $d$ in the range of 4.6 to 4.8 nm are significantly less populated in glycosylated GluN1 LBD in comparison to non-glycosylated GluN1 LBD (percentile bootstrap, confidence level of 95%, see section S3; for the exact definition of $d$, see Fig. 2a and section S5).

Glycine binding promotes the closing of the GluN1 LBD.[18,20,23] Our simulations show that the effect of glycosylation is similar to the effect of coagonist binding. However, glycosylation alone, in the absence of glycine, is insufficient to change the population of closed forms of GluN1 LBD to 100%. We predict that glycosylation potentiates the closure of GluN1 LBD by a coagonist. The effect of glycosylation on the structure of LBDs is difficult to deduce from currently available experimental structures of GluN1 LBD, because most of the X-ray structures refer to non-glycosylated proteins, while only three[39,40] refer to partially glycosylated proteins and do not fully resolve the glycans. MD simulations provide a detailed dynamic model of glycans in GluN1 LBD.

The results for the GluN2B simulations mirror the results for the GluN1 monomer, albeit to a more modest but significant degree. Despite starting from just one structure (mainly based on PDB entry 4PE5; see more details in section S5), the GluN2B LBD populates a wide distribution of conformations. Furthermore, our GluN2B simulations show that glycosylation results in a distribution of conformations that are skewed more toward closed-like states when compared with the non-glycosylated form (Fig. 2e). The difference between the probability distribution functions is significant in the ranges of the interlobe distances of 3.6 to 3.9 nm, where the glycosylated form is more stable, and 4.1 to 4.7 nm, where the non-glycosylated form is more stable (percentile bootstrap, confidence level of 95%, see section S3). These results further our observations that glycosylation leads the LBDs to a more closed conformation, which may affect the activity of the ion channel.

**Simulations predict a mechanism of the potentiating effect of glycosylation.** In closed conformations, the Man5 moiety at residue GluN1-N440 (located in the disordered region between β-sheets 3 and 4, in the terminology of ref. 23) in the lobe S1 of the GluN1 LBD can approach the lobe S2 of GluN1 LBD in the region of residues 710 to 723 (helix H and the disordered region between helices G and H), allowing for noncovalent interactions between the two lobes (Fig. 3a). On the protein side, these interactions involve terminal oxygen atoms from the side chains of residues Glu712, Glu716, Gln719 and/or Asp723. On the side of the glycan, most OH groups from Man5 can transiently participate in the interactions, resulting in numerous conformations of the formed complex, without a single preferred structure (Fig. 3c). Though we have not performed simulations of GluN1 LBD with glycans other than Man5 due to the high computational cost of such simulations, the nonspecific involvement of the hydroxyl groups from the Man5 glycan in the interlobe interactions suggests that other glycan types on residue N440 in the GluN1 LBD may lead to a similar effect.

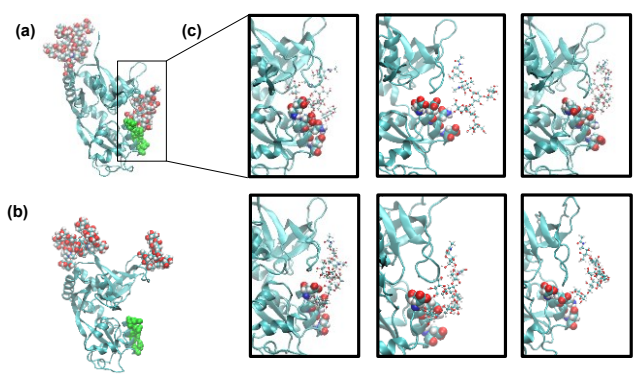

**Fig. 3.** (a) Man5 glycan at residue GluN1-N440 (*van-der-Waals spheres, right*) and amino acid residues Glu712, Glu716, Gln719 and Asp723 from the other lobe of GluN1 LBD (*green*) transiently noncovalently interact, which explains why closed conformations of the GluN1 LBD are more stable in the glycosylated state. (b) In open conformations of the GluN1 LBD, the Man5 glycan and the other lobe of the glycoprotein do not interact. (c) Representative structures for the transient interactions between Man5 glycan (*CPK representation*) and residues 712, 716, 719, 723 (*van-der-Waals spheres*) illustrate that no single stable structure exists under physiological conditions.

In open conformations, Man5 and the lobe S2 of glycosylated GluN1 LBD are too far from each other to interact (Fig. 3b). This conclusion is evident from the analysis of the two-dimensional distribution between (1) the distance $d$ between $C_\alpha$ atoms in residues 507 and 701, quantifying how open or closed a current conformation of GluN1 LBD is (Fig. 2a), and (2) the distance $d_{g\text{-}ol}$, defined as the shortest distance between heavy atoms in the glycan attached to N440 and heavy atoms in the other lobe of the protein (residues 710 to 723), quantifying whether the glycan interacts with the other lobe of the protein or not. This two-dimensional distribution across all 536,651 snapshots in all 262 MD trajectories of glycosylated GluN1 LBD is shown in Fig. 4. The empty field in the lower right part of the plot shows that the interactions between the glycan and the lobe S2 do not occur in all open-clamshell conformations sampled in our simulations.

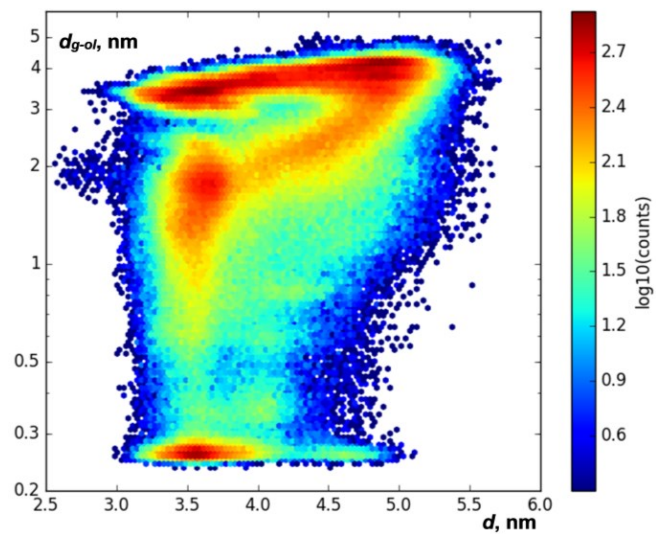

**Fig. 4.** Glycan attached to GluN1-N440 interacts with the opposite lobe of the protein only in closed-clamshell conformations of GluN1 LBD. This heatmap shows a two-dimensional distribution of geometries of glycosylated GluN1 LBD in all 536,651 frames of 262 MD trajectories in terms of two variables: $d$, measuring whether a conformation is clamshell open/closed, and $d_{g\text{-}ol}$, the shortest distance between heavy atoms in the glycan attached to N440 and heavy atoms in the other lobe of the protein (residues 710 to 723). (Note the log scales on the $y$ axis and the colorbar). The empty field in the region of the diagram with $d > 5.2$ nm and $d_{g\text{-}ol} < 0.5$ nm implies that no open-clamshell conformations with the glycan interacting with the opposite lobe of the protein have been reached.

Similar interactions were seen in simulations of the GluN2B LBD, with some notable differences. The GluN2B-N688 glycan is bound to the beginning of helix F, close to the hinge between the lobes S1 and S2 of the GluN2B LBD and close to where glutamate binds. The glycan at this position is seen to interact extensively with residues on Loop 2, β-sheet 6, and helix D. Specifically, it appears that the hydroxyls of the glycan interact with residues Glu518, Lys488, Lys489, His486, Trp494, Glu517, and Arg519 (from most heavily engaged to least). However, unlike in GluN1, it appears the other two glycans may also potentiate closure of the clamshell. The glycan at GluN2B-N444, located between β-sheets 3 and 4, also interacts with the lobe S2 at the start of helix H, in particular with residues Arg712, Asp715, Asp716. These residues are similar to those we have found in GluN1. Further, the GluN2B-N491 glycan is bound to Loop 2 between β-sheets 6 and 7, and while it does not interact significantly with the lobe S2, it is close enough to the GluN2B-N688 that they interact in 15% of the frames in our simulations. GluN2B-N444 also interacts with GluN2B-N688, though only in 6% of the frames (data not shown). Therefore, it appears that all of the glycans on GluN2B LBD could also potentiate closure, with GluN2B-N688 likely being the most influential.

In our simulations, GluN2B does not sample open conformations as widely as those seen in the GluN1 domain. This may be due to the fact that the GluN2B simulations have started from a single closed conformation and have yet to explore those fully open conformations. However, in contrast to the GluN1 glycans, which are linked to the protein at residues far away from the opposite lobe, two of the GluN2B glycans (N491, N688) are linked to residues that are relatively close to the opposite lobe and one another.

**Table 1**. Timescale of the opening/closing transition in the glycosylated GluN1 LBD and GluN2B LBD are on the order of 0.5 and 0.2 μs, respectively. Surprisingly, adding glycans to GluN1 and GluN2B LBDs does not affect the rate of opening/closing, unlike the relative stability of the open-clamshell and closed-clamshell conformations.

| Method used for estimate | Timescale, μs | | Ratio of time-scales |
|---|---|---|---|
| | with glycans | without glycans | |
| GluN1 LBD | | | |
| MSM, 99 clusters, lag 256 ns | 0.52 | 0.53 | 0.99 |
| MSM, 6 clusters, lag 256 ns | 0.49 | 0.50 | 0.99 |
| MSM, 99 clusters, lag 128 ns | 0.40 | 0.38 | 1.04 |
| tICA | 0.42 | 0.43 | 0.98 |
| GluN2B LBD | | | |
| MSM, 99 clusters, lag 256 ns | 0.21 | 0.19 | 1.14 |
| MSM, 6 clusters, lag 256 ns | 0.16 | 0.17 | 1.11 |
| MSM, 99 clusters, lag 128 ns | 0.13 | 0.12 | 1.08 |
| tICA | 0.18 | 0.18 | 1.00 |

**Kinetics of clamshell-like opening and closing of GluN1 and GluN2B LBDs.**

GluN1 and GluN2B LBD opening/closing, according to our simulations, are fast and occur on the sub-microsecond timescales. To quantitatively determine the timescales from our MD trajectories, we used Markov state models (MSMs), an approach successfully applied in the past to other biomolecular systems.[30-32] For the glycosylated form, the MSM with the optimal number of clusters (99 clusters, see section S1) predicts the slowest timescales of the opening/closing motion to be 0.5 and 0.2 μs in the GluN1 and GluN2B subunits, respectively, and some other plausible models yield comparable timescales (Table 1). The slowest timescales reported in Table 1 are aggregate characteristics of complex transitions between a continuous spectrum of more open and more closed conformations of the LBDs. To a first approximation, the slowest timescale is comparable, by an order of magnitude, to the typical timescale of opening the clamshell of the LBD, as well as the typical timescale of its closing (see section S4). With the available amount of sampling, we have not found any significant difference in the timescale of opening/closing transitions in glycosylated and non-glycosylated LBDs (Table 1). To the best of our knowledge, no experimental data on NMDAR LBDs opening/closing on the sub-microsecond timescale have been published so far. The results on GluN1 LBD dynamics obtained by single molecule fluorescence resonance energy transfer (smFRET)[41-43] refer to millisecond timescales, and hence they can not be compared to the results of our simulations that make predictions on the microsecond timescale. No similar data on GluN2B LBD dynamics, to the best of our knowledge, have been published.

Opening and closing of NMDARs as ion channels occur mainly on the timescales of 0.1 to 100 ms,[44] two to five orders of magnitude slower than the timescales of clamshell-like opening and closing of LBDs that we predict in this work. Therefore, there should be no direct mechanical coupling between each event of a conformational change in LBDs, and opening or closing of the ion channel pore. Instead, the functional state of the full receptor (open ion channel, closed ion channel, etc.) must be controlled by changes in time-averaged populations of various functional states of LBDs (open-clamshell vs. closed-clamshell). Discussion of mechanisms of interactions between GluN1 and GluN2B LBDs and the other parts of NMDARs[33,34] goes beyond the scope of this work.

**Experimental validation of the potentiating role of glycans at GluN1 LBD in full-length NMDARs.** Our molecular dynamics simulations suggest that glycosylation on GluN1-N440 stabilizes the activated (closed clamshell) structural state of the GluN1 LBD. This stabilization might enhance the ability of glycine to bind to the GluN1 LBD. Therefore, selective deglycosylation of GluN1-N440 might be expected to increase the EC50 for glycine. In order to test this prediction, we introduced the mutation N440Q into GluN1 by site-directed mutagenesis, which is a standard technique to completely prevent attachment of the N-linked glycan to this residue.[7,13] We expressed wild-type (WT) and mutated full-length NMDARs in *Xenopus* oocytes and measured the glycine EC50 in voltage-clamp electrophysiology experiments (Fig. 5a,b). Normalized dose-response curves were fitted using a standard Hill function (Fig. 5c).

Whereas WT GluN2A-GluN1 channels were found to have glycine EC50 = 2.28 ± 0.03 μM, mutant GluN2A-GluN1 (N440Q) channels have glycine EC50 = 3.43 ± 0.05 μM, a statistically significant 50% increase in the glycine EC50 ($p < 0.01$, Student's t-test). Thus, while the glycan attached to GluN1-N440 does not open NMDAR channels directly, it enhances the ability of glycine to activate the channels (in the presence of glutamate).

Since glycine is an amino acid naturally present *in vivo*, the possibility of background glycine contamination should be taken into account. In our *Xenopus* oocyte assays, this contamination was negligible. We did not observe any current in the absence of glycine even at saturating glutamate concentrations. At the glycine concentration of 300 nM we observed noticeable NMDAR currents, indicating that the concentration of contaminating glycine, if present, must have been lower than 300 nM. Possible trace amounts of contaminating glycine (on the order of 100 nM) should have been present to the same extent in the case of both WT and mutated NMDARs experiments, and therefore could not account for the shift in the glycine EC50 that we observed.

## DISCUSSION

Our work for the first time puts forward a hypothesis about the physiological role of specific glycans on NMDARs, specifically glycans at GluN1 N440 and GluN2B N688 residues, and supports it with experimental data. We find that intramolecular interactions involving glycans at the above-mentioned residues affect the conformations of the corresponding LBDs similarly to binding agonists, stabilizing closed-clamshell conformations of the LBDs. This suggests that the glycans at GluN1-N440 and GluN2B-N688 play intramolecular potentiator roles in NMDARs.

The novelty of the present work is in the use of computational modeling to address the consequences of abnormal NMDAR glycosylation. The advantages of the computational approach include a unique degree of control over the model of the system under investigation and its detailed description on the atomic resolution level. In particular, we set up simulations such that they refer to physiological conditions (temperature of 310 K, water solution with the physiological salt concentration, primary structure exactly as in the human NMDARs, no interactions between different copies of glycoproteins, and the glycosylation pattern as *in vivo*). We recorded coordinates of each atom in the system every 0.2 ns. Though the computational approach provides a model-dependent information about the system and cannot replace experiments, it can yield models with detailed structural information about the system under investigation under physiological conditions that can rationally guide further experiments.

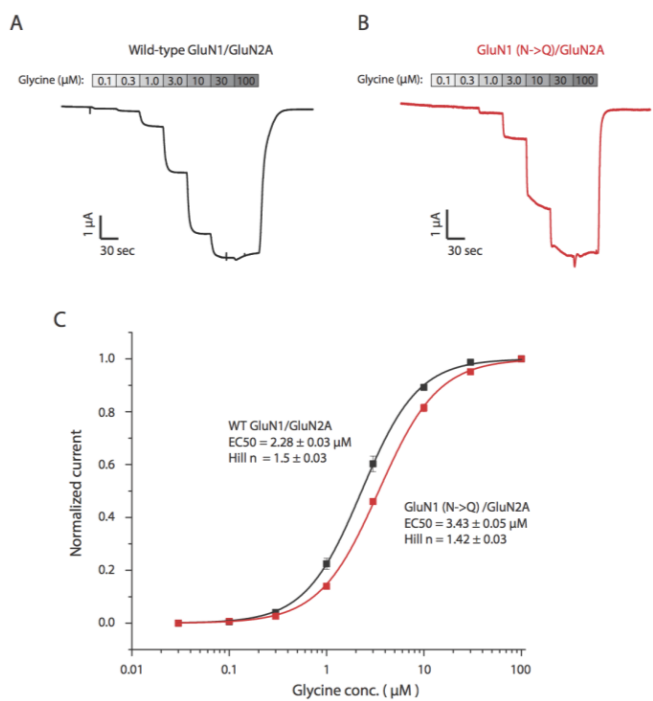

**Fig. 5.** The mutation GluN1-N440Q in the GluN1/GluN2A NMDA receptor results in a small but detectible rightward shift in the glycine EC50. (a,b) The wild-type and N-to-Q mutant channels were expressed in oocytes and the dose response was measured using two-electrode voltage clamp recordings. (c) Averaged data from $n$ = 12 recordings of each were plotted and fit with a Hill function to reveal a 50% increase in the glycine EC50 in the presence of the N440Q mutation, which makes glycosylation at this residue impossible.

On the other hand, computational modeling has certain limitations, including the size of systems that can be studied, the timescales of captured processes, and distortions introduced by employed models. We had to limit the size of the simulated system to GluN1 or GluN2B LBDs of NMDAR. This approximation is justified by the experimentally established modular nature of NMDARs.[15,16] Nevertheless, expanding simulations to full NMDARs and complexes of NMDARs with other proteins is desired in the future, but would be an immense computational undertaking (with up-to-date computational facilities, it would take ~5 years to carry out similar simulations for the full NMDAR). Our simulations revealed dynamic processes in the GluN1 and GluN2B LBDs on the timescales below 1 μs. For comparison, Markov state models typically capture timescales comparable to the aggregate duration of all used MD trajectories (in this work, 0.1 – 0.3 ms for each system). Therefore, any processes on the timescale of ~1-100 μs are unlikely to occur in our simulations of GluN1 and GluN2B LBDs. This work does not attempt to address dynamics on the millisecond and longer timescales.[41-43] Finally, the degree of uncertainty introduced by the use of specific models can be estimated from Table 1, which shows that the timescales of opening/closing in GluN1 and GluN2B LBDs predicted by several models are qualitatively the same.

We experimentally confirmed the potentiating role of the glycan attached to GluN1-N440, predicted from our computations, by measuring changes in glycine EC50 after glycan removal using the GluN1-N440Q point mutation in the full receptor. Though our prediction on the role of this glycan was made based on computer simulations of a separate GluN1 LBD, and in principle could appear to be not transferable to the full receptor, the experimental validation was performed for the full-length NMDAR, demonstrating the transferability of the potentiating role of the glycan. Our observation that the glycine EC50 increases by approximately 50% is consistent with the idea that the N440 glycan stabilizes the glycine-bound state of GluN1 in the wild-type NMDA receptor. To estimate the scale of possible biomedical implications of such a change in the glycine EC50, consider the following data. Mutation L812M in the GluN2A subunit changes the glycine EC50 by a factor of 3.6 (in the receptor with one wild-type and one mutated GluN2A subunits) or 12 (with both GluN2A subunits mutated), and the glutamate EC50 by 4 and 10, respectively. A patient with this mutation had profound global developmental delay with no attainment of any milestones since birth.[45] Mutation V667I in GluN2D changes EC50 for glycine and glutamate by factors of 1.7 and 1.5, respectively. This mutation was found in patients with epileptic encephalopathy and global developmental delay.[46] Mutations G815R and F817L in GluN1 change EC50 for glutamate by factors of 4.4 and 4.2, respectively (data for glycine not reported). The phenotypes included severe intellectual disability, movement disorder, seizures.[47] Multiple experimental data show that the coagonist binding sites on NMDARs are not 100% saturated *in vivo*, at least in some important locations in the brain (e.g., in hippocampus or prefrontal cortex),[48-50] therefore, changes in the coagonist EC50 may lead to dramatic changes in the NMDA-dependent currents. Thus, it is not unreasonable to expect that the change in the glycine EC50 by a factor of 1.5 that we report might lead to neurological disorders, though maybe less severe than those listed above.

Most structural studies of NMDARs (or their fragments) investigate the glycan-free forms.[20,23-27] The absence of glycans in the GluN1 LBD structures in the cited works was possibly a side effect of using *E. coli* as the expression system. Likewise, past publications that undertook computational modeling of NMDARs or their parts[19-22,33-37] have also omitted glycans. However, the absence or presence of glycans may significantly change functionally relevant conformations, as our results on Man5 at GluN1-N440 and GluN2B-N688 suggest.

We also make a number of testable predictions in this paper, in addition to the change in glycine affinity mentioned above, namely:

1) The clamshell-like opening/closing motions of GluN1 and GluN2B LBDs occur on a sub-microsecond timescale, which could be checked, for example, by T-jump IR spectroscopy or electron paramagnetic resonance (EPR). The previous literature does not seem to be consistent on the timescale of these motions. On the one hand, no significant free energy barrier between the clamshell-open and closed states was found in previous simulations of GluN1 LBD.[20] On the other hand, the opening/closing motion was reported to occur on the millisecond timescale, as determined by smFRET, a method with millisecond temporal resolution.[41-43] Our prediction suggests using sub-microsecond-resolution methods to investigate the opening/closing motion of GluN1 and GluN2B LBDs such that it is decoupled from other conformational transitions that may occur in GluN1 and GluN2B LBDs on the millisecond timescale.

2) The equilibrium concentrations of clamshell-closed conformations after glycosylating (de-glycosylating) residue N440 in the separate GluN1 LBD module and, presumably, in the full NMDAR, increase (decrease, respectively); the same for (de)glycosylation at residue N688 in GluN2B subunit.

3) The changes in the equilibrium concentrations of clamshell-closed conformations of GluN1 LBD are disrupted by mutations in the residues involved in the transient interactions with Man5 glycan at N440 (Glu712, Glu716, Gln719 and Asp723 in GluN1 subunit). Because of the non-specific character of interactions between the glycan and these four amino acid residues, mutations in all (or maybe most) of them are required for a noticeable effect on the relative stability of the GluN1 LBD conformations.

All these predictions, to the best of our knowledge, have not been experimentally tested so far.

The identification of NMDAR glycosylation as important for agonist affinity could have potential medical consequences if glycosylation were differentially regulated under physiological and pathological situations. As NMDAR dysfunction is implicated in diseases including autism, epilepsy and schizophrenia, our work suggests that future studies could look for abnormal NMDAR glycosylation, especially at GluN1 N440 and GluN2B N688 residues, in the studies of various neurological disorders.

## METHODS

**Molecular dynamics simulations.** We ran all-atom MD simulations of glycosylated and non-glycosylated GluN1 LBDs (Fig. 1b-c) in explicit solvent under physiological conditions (310 K, water solution with the ion strength of 0.154 M, amino acid sequence exactly as in the human NMDARs (uniprot code Q05586-3), glycosylation with Man5 at residues 440, 471 and 771). Simulations started from 23 different geometries corresponding to different functional states of the GluN1 LBD. In total, 262 and 196 MD trajectories with the aggregate simulation time of 0.107 ms and 0.106 ms were generated for the glycosylated and non-glycosylated GluN1 LBDs, respectively. We followed up these simulations with similar ones on GluN2B LBD. The GluN2B subunit is known to be glycosylated at residues 444, 491, and 688. All-atom simulations of the glycosylated and non-glycosylated forms of GluN2B were performed using a model built from the full channel structures. This resulted in 247 and 613 trajectories totaling 0.086 and 0.344 milliseconds for the glycosylated and non-glycoslated forms, respectively. For more details, see section S5.

**Interpretation of molecular dynamics trajectories.** Markov state models[30] were built to reconstruct the thermodynamic and kinetic properties of the NPT ensembles of glycosylated and non-glycosylated GluN1 LBD and GluN2B LBD proteins at equilibrium from finite-length MD trajectories, each of which did not completely sample the configuration space. Markov state models with 99 clusters and the lag time of 256 ns were used [see Supplementary Information (SI), sections S1 and S2]. For more details, see section S5.

**Glycine EC50 measurements.** Expression of NMDAR channels in *Xenopus* oocytes was achieved by subcloning the human cDNAs for these channels into the pTNT vector (Clontech). We expressed GluN1 paired with GluN2A, which provides rapid and efficient expression of the full-length NDMARs in *Xenopus* oocytes. Mutagenesis was carried out using the Quikchange Lightning Multi kit (Agilent) as per manufacturer's instructions. To produce RNA for injection into oocytes, the constructs were linearized, purified, and used as the substrate for T7 RNA polymerase-mediated RNA synthesis (mMessage mMachine T7, Ambion). Oocytes were de-folliculated with 2 mg/ml collagenase type 2 in OR-2 solution (in mM: 82.5 NaCl, 2.4 KCl, 1 $MgCl_2$ and 5 HEPES), and then transferred to ND-96 solution (in mM: 96 NaCl, 2 KCl, 1 $MgCl_2$, 1.8 $CaCl_2$ and 5 HEPES). The N440Q mutant GluN2A-GluN1 channels expressed at similar levels as wild-type GluN2A-GluN1 channels. For two-electrode voltage-clamp (TEVC) recordings, oocytes were injected with mRNA for hGluN1-1a and hGluN2A, using either WT or mutant mRNA. Glycine dose response experiments were performed by perfusing increasing concentrations of glycine (from 0.1 μM to 100 μM) onto the oocytes (all solutions contained 100 μM glutamate to allow glycine-dependent NMDAR activation). Measurements were made on both wild-type and mutant channels in side-by-side recordings to reduce variability. The N-to-Q mutation is a common technique used to prevent glycosylation of proteins at specific positions. It has previously been successfully applied to NMDARs.[7,13] We have not been able to make measurements of EC50 for the NMDAR with mutation GluN2B-N688Q, because the protein with this mutation did not express for unknown reasons.

## ACKNOWLEGEMENTS
The authors acknowledge support from NIH (grant R01GM062868-14). We are grateful to Ariana Peck, Matthew Harrigan and Keri McKiernan for helpful comments on the manuscript. We thank the users of the Folding@home distributed computing project who donated compute time used for the simulations. This work also used the XStream computational resource, supported by the National Science Foundation Major Research Instrumentation program (ACI-1428930), and Sherlock cluster. We would like to thank Stanford University and the Stanford Research Computing Center for providing computational resources and support that have contributed to these research results.

## AUTHOR CONTRIBUTIONS
A.V.S. conceived and designed the project, with supervision from V.S.P. A.V.S. and N.H.S. carried out simulations. D.H.H. and J.E.H. designed and conducted experiments. A.V.S. wrote the paper, with important contributions from N.H.S. and D.H.H. All co-authors interpreted the results and edited the paper. V.S.P. and B.D.S. supervised the project and provided advice.

# Computationally Discovered Potentiating Role of Glycans on NMDA Receptors

*Anton V. Sinitskiy,*[1] *Nathaniel H. Stanley,*[2,3] *David H. Hackos,*[4] *Jesse E. Hanson,*[4] *Benjamin D. Sellers,*[3] *& Vijay S. Pande*[1,5]

[1] Department of Chemistry, Stanford University, Stanford, California 94305

[2] Stanford ChEM-H, Stanford University, Stanford, California 94305

[3] Department of Discovery Chemistry, Genentech, Inc., 1 DNA Way, South San Francisco, CA 94080, USA

[4] Department of Neuroscience, Genentech, Inc., 1 DNA Way, South San Francisco, CA 94080, USA

[5] Department of Computer Science and Department of Structural Biology, Stanford University, Stanford, California 94305

Correspondence and requests for materials should be addressed to A.V.S. (email: sinitskiy@stanford.edu) and V.S.P. (email: pande@stanford.edu).

# SUPPLEMENTARY INFORMATION

## Section S1. Choice of the optimal number of Markov states

The optimal number of Markov states in the Markov state model of the glycosylated GluN1 LBD was chosen by cross-validation with the use of generalized matrix Rayleigh quotient (GMRQ)[1] for the slowest implicit timescale as the score function. 25 various random divisions of the dataset of all MD trajectories for glycosylated GluN1 LBD into a training set (80% of trajectories) and a test set (20% of trajectories) were performed. Markov state models with the number of Markov states ranging from 2 to 1000 were built from each training set, and then the performance of each MSM was scored against the corresponding test set. Scores averaged over 25 various divisions are reported in Fig. S1a. As expected, the average score for training sets increases with the increase of the number of states, while the average score for test sets reaches a maximum value and decreases with the further increase of the number of states because of overfitting. The optimal number of states corresponding to the maximum average test score turns out to be 99. As for the glycosylated GluN2B LBD, a similar procedure leads to the optimum number of Markov states maximizing the average test score to be 118 (Fig. S1b). In both cases, the score is not very sensitive to the number of states in the vicinities of the extrema, allowing us to expect reasonable performance of Markov models in a wide range of the numbers of Markov states, from ~50 to ~500. Taking this lack of sensitivity into account, we used the same number of Markov states, namely 99, to model all four molecular systems under investigation, glycosylated and non-glycosylated GluN1 and GluN2B LBDs, to ensure the comparability of the results for glycosylated and non-glycosylated forms, as well as different subunits.

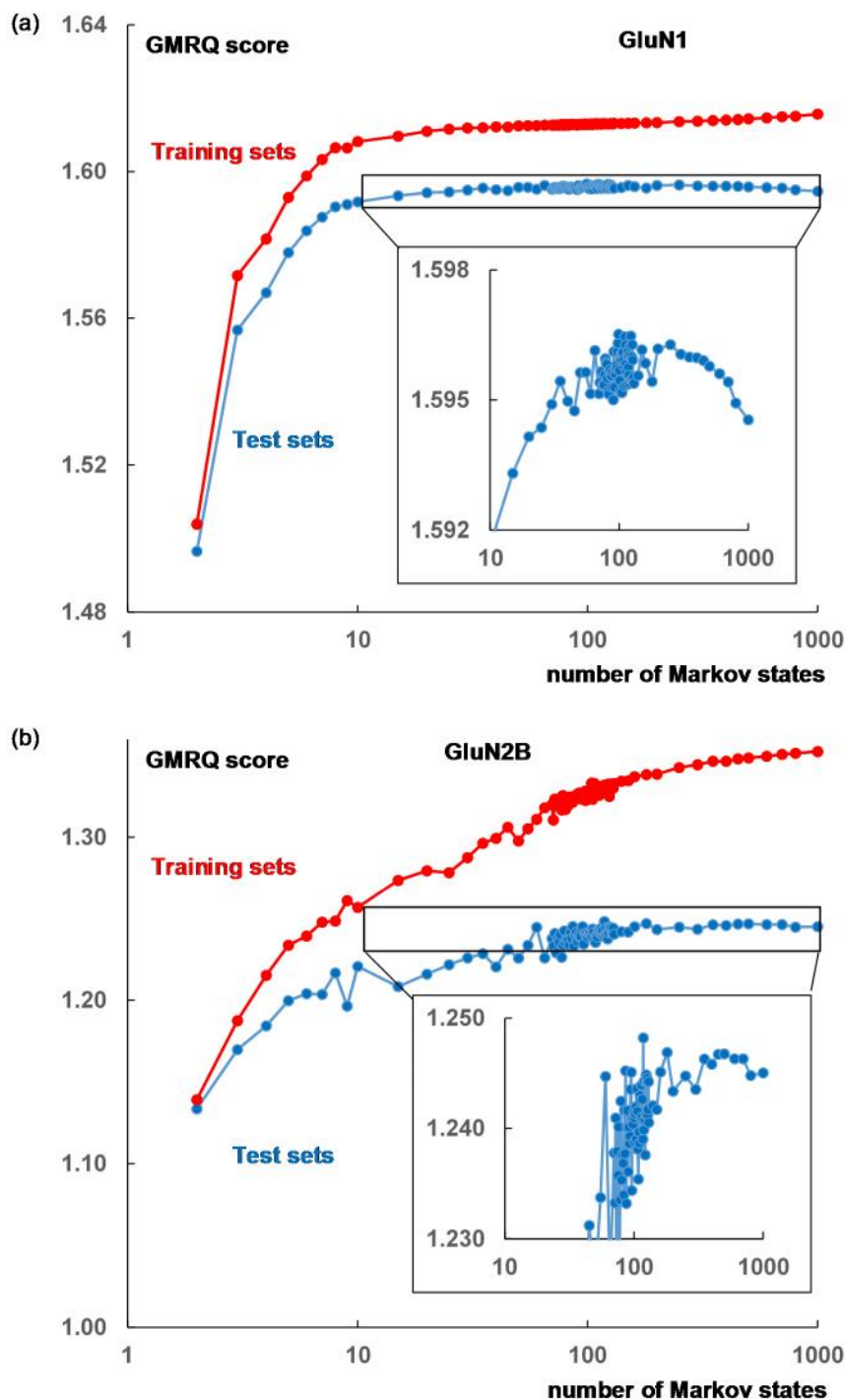

**Fig. S1.** (a) Cross-validation with the use of generalized matrix Rayleigh quotient (GMRQ) for the slowest implicit timescale as the score function was used to choose the optimal number of Markov states to model glycosylated GluN1 LBD, namely 99. Average scores for training sets (*red*) and test sets (*blue*) over 25 random divisions of the data set into training and test sets are provided. (b) In the case of our glycosylated GluN2B LBD simulations, the optimal number of Markov states appears to be 118, though the model with 99 states is relatively good, too. For comparison, we used Markov models with 99 states both for GluN1 and GluN2B LBDs.

## Section S2. Choice of the optimal lag time

Lag time is the distance in time between neighboring frames from MD trajectories used to estimate the MSM transition matrix. Too small values of the lag time can lead to inadequacy of MSM due to a violation of the Markov assumption at short timescales. Too large values of the lag time lead to poorer statistics on transitions. Analysis of the convergence of implicit timescales of MSMs as a function of lag time (Fig. S2) can be used to choose the optimal value of lag time. For our simulations of glycosylated GluN1 LBD, we chose the lag time of 256 ns. In the case of GluN2B LBD, the overall picture is the same as in GluN1 LBD. For the purpose of comparison, this lag time was used to model all four molecular systems under investigation, glycosylated and non-glycosylated GluN1 and GluN2B LBDs.

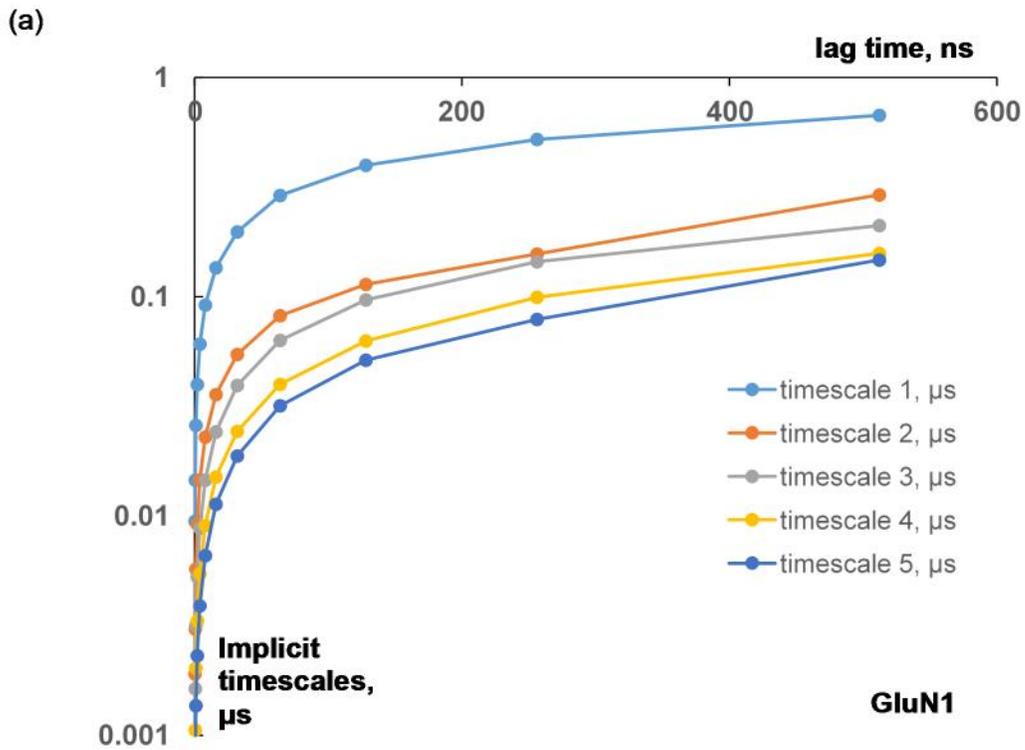

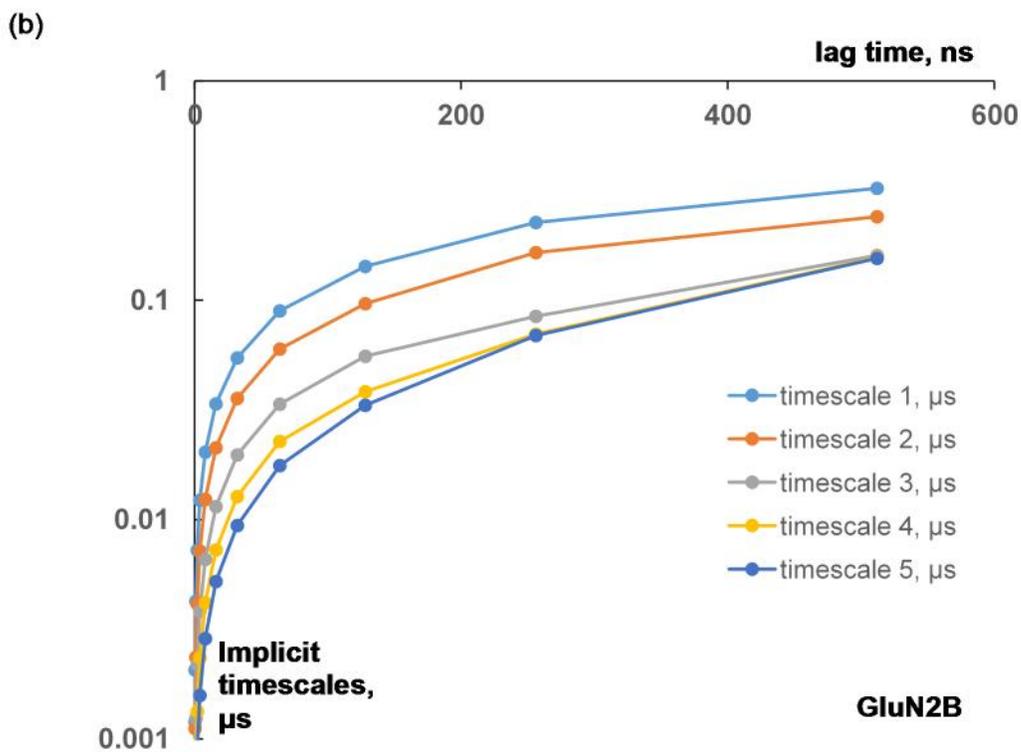

**Fig. S2**. (a) Five slowest implicit timescales, estimated from MSM for glycosylated GluN1 LBD with 99 states, converge to stationary values as the lag time increases, confirming the applicability of the Markov assumption. We chose the lag time of 256 ns as a compromise between ensuring Markov and good statistics on transitions between Markov states. Note the gap between the first and the second slowest timescales by a factor of 2.5-4 (depending on the lag time), implying a relatively good decoupling of clamshell opening/closing motion from other conformational transitions in glycosylated GluN1 LBD. (b) In GluN2B LBD, the overall picture with implicit timescales is the same. However, the timescales are faster (up to ~0.2 μs, vs. up to ~0.5 μs in GluN1), and the gap between the first and the second slowest timescales is smaller (a factor of ~1.4-1.6).

## Section S3. Estimate of the statistical significance of the difference between probability distribution functions for glycosylated and non-glycosylated GluN1 and GluN2B LBDs by bootstrapping

We performed 1000 rounds of resampling of time series $d(t)$ for glycosylated, and, separately, non-glycosylated GluN1 LBD. For each resampled data set, the probability distribution functions for glycosylated and non-glycosylated GluN1 LBD, denoted below as $f_1(d)$ and $f_2(d)$, respectively, were computed with the use of Markov state models with 99 states and the lag time of 256 ns. Then, the difference between the two probability distributions, $df(d) = f_1(d) - f_2(d)$, was computed for each of 1000 sets of resampled data. Confidence intervals for $df(d)$ were computed at each $d$ by percentile bootstrap of 1000 estimates of $df(d)$. The values of $d$ for which the probability distribution functions for glycosylated GluN1 LBD statistically significantly exceeds the probability distribution functions for non-glycosylated GluN1 LBD were determined as those values of $d$ at which the confidence intervals for $df(d)$ are entirely above zero, and *vice versa*. The confidence intervals for $df(d)$ for the P-values of 95% and 99% are shown in Fig. S3a. Similar computations were performed for GluN2B LBD (Fig. S3b).

(see next page)

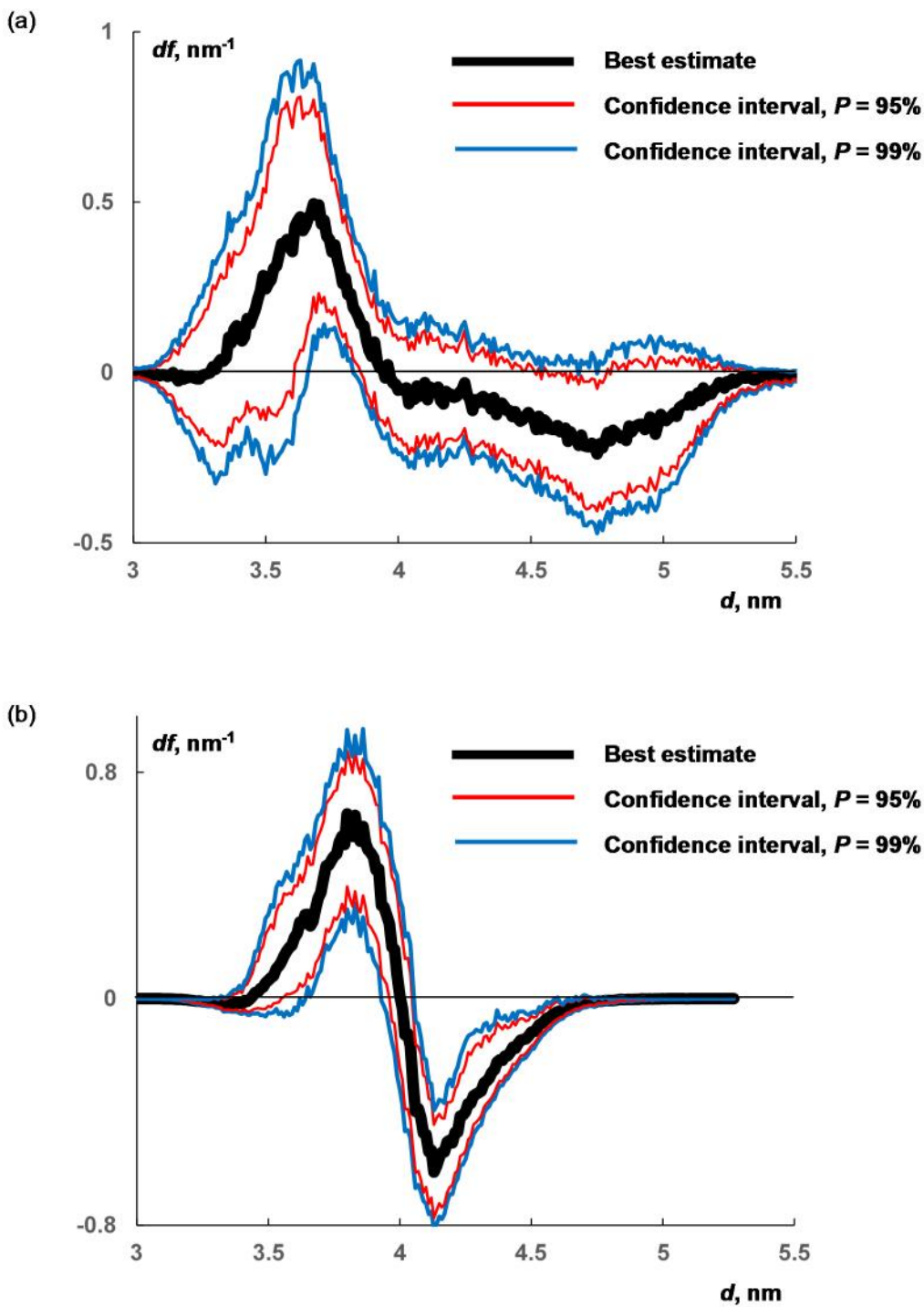

**Fig. S3.** (a) Conformations of GluN1 LBD with *d* in the range of 3.61 to 3.85 nm are statistically significantly more populated (the confidence interval for *df*, which is the difference between probability distribution functions for glycosylated and non-glycosylated GluN1 LBD, is above zero at these values of *d*), and those with *d* in the range of 4.57 to 4.78 nm are statistically significantly less populated (the confidence interval for *df* is below zero) in glycosylated GluN1 LBD than in non-glycosylated GluN1 LBD (percentile bootstrap, confidence level of 95%).

*Black*: the best estimate of *df* based on the available MD trajectories (in other words, the red curve from Fig. 2c minus the blue curve from Fig. 2c). *Red*: the confidence interval for *df* with the confidence level of 95% estimated by percentile bootstrap. *Blue*: the same, with 99% confidence level. (b) In GluN2B LBD, the glycosylated form is statistically significantly more stable in the ranges of 3.55 to 3.95 nm, and less stable between 4.05 to 4.69 nm (percentile bootstrap, confidence level of 95%).

## Section S4. Physical interpretation of slowest timescales of Markov state models

To clarify the meaning of the slowest Markov state model timescale, consider a simplified physico-chemical two-state model of GluN1 or GluN2B LBD, with one open and one closed states (*O* and *C*, respectively),

$$O \underset{k_{op}}{\overset{k_{cl}}{\rightleftharpoons}} C, \qquad (1)$$

and assume that the transitions between these two states are described by the first order reaction kinetics, with the rate constants $k_{cl}$ and $k_{op}$ for closing and opening, respectively. Then the concentration of the open and closed forms will depend on time *t* as follows:

$$[O] = c_0 + c_1 e^{-(k_{cl}+k_{op})t}, \qquad [C] = c_2 + c_3 e^{-(k_{cl}+k_{op})t}, \qquad (2)$$

where constants $c_0$, $c_1$, $c_2$ and $c_3$ are determined by the initial concentrations of the open and closed forms and the ratio of the rate constants $k_{cl} / k_{op}$. On the other hand, the time evolution of the ensemble of Markov processes for an *n*-state Markov chain can be written in comparable notations as

$$[O] = c'_0 + \sum_{i=1}^{n-1} c'_i e^{-t/\tau_i}, \qquad [C] = c''_0 + \sum_{i=1}^{n-1} c''_i e^{-t/\tau_i}, \qquad (3)$$

where $\tau_i$ are characteristic timescales of the Markov chain sorted in the descending order, $\tau_1 \geq \tau_2 \geq ... \geq \tau_{n-1}$. If the slowest characteristic timescale $\tau_1$ is significantly greater than the other characteristic timescales, then for sufficiently large *t* eq. (2) approximates eq. (3) with

$$\tau_1 = \frac{1}{k_{cl}+k_{op}} = \frac{\ln 2}{\frac{1}{\tau_{cl}} + \frac{1}{\tau_{op}}}, \qquad (4)$$

where $\tau_{cl}$ is the time over which half of all open conformations would close, provided that closed conformations could not open, and $\tau_{op}$ is the time over which half of all closed conformations would open, provided that open conformations could not close. The equilibrium concentrations of the open and closed forms of GluN1 or GluN2B LBDs are comparable by the order of magnitude, as Fig. 2(c,e) shows. Therefore $k_{cl} \sim k_{op}$, and hence, $\tau_{cl} \sim \tau_1, \tau_{op} \sim \tau_1$. Thus, the slowest timescale reported in Table 1 is an aggregate characteristic of complex transitions between more open and more closed conformations of the GluN1 and GluN2B LBDs. In the first approximation the slowest timescale is comparable, by the order of magnitude, to typical timescales of both opening and closing the clamshell.

## Section S5. Methods: more detailed description

**Molecular dynamics simulations.** Initial 23 geometries of GluN1 LBD were prepared based on all experimental X-ray structures of GluN1 LBD available at the moment when our simulations started (PDB codes: 1PB7, 1PB8, 1PB9, 1PBQ,[2] 1Y1M, 1Y1Z, 1Y20,[3] 2A5T,[4] 4KCC,[5] 4KFQ,[6] 4NF4, 4NF5, 4NF6, 4NF8,[7] 4PE5,[8] 4TLL, 4TLM[9]). The proteins consisted of residues 393 to 546 and 663 to 800 (residue numbering everywhere in this article corresponds to the full GluN1 sequence). The primary structure of GluN1 subunit from *H. sapiens*, isoform 3 (NR1-3), uniprot identifier Q05586-1, was used. The structures were solvated in TIP3P water with sodium and chloride ions in the amount corresponding to the physiological concentration of 0.154 M. Glycosylation was performed with the use of Glycoprotein Builder.[10] Energy minimization, heating and two-stage preequilibration resulted in preequilibrated structures used for production simulations. These preequilibrated structures were within 1 Å from the original X-ray structures in terms of RMSD for the protein backbone atoms.

For the GluN2B simulations, a homology model of the LBD was built from full NMDAR X-ray structures (PDB codes: 4PE5,[8] 4TLL, 4TLM[9]). The six LBD domains (two from each structure) were aligned and used to build a consensus model using Schrodinger's Maestro software (version 2015-1).[11] The template sequence consisted of residues 401 to 604 and 658 to 806, with a glycine linker between 604 and 658. The sequence was based off of the GluN2B subunit from *H. sapiens*, uniprot identifier Q13224-1. The structures were solvated in TIP3P water with sodium and chlorine ions added to neutralize the system. For the glycosylated system, glycosylation was performed with the use of the Glycoprotein Builder.[10]

MD simulations were performed with Amber ff99SB-ILDN[12] and GLYCAM_06i[13] force fields (for the protein and glycan parts, respectively), resulting in 262 MD trajectories for glycosylated and 196 MD trajectories for non-glycosylated GluN1 LBD, with the aggregate duration of 107 and 106 μs, respectively. For the GluN2B LBD, 247 trajectories for the glycosylated and 613 trajectories for the glycosylated form were performed, resulting in an aggregate simulation time of 86 and 344 μs, respectively. Simulations were run on Folding@home and various types of GPUs available on Stanford computer clusters (Sherlock, XStream). Depending on the type of the used GPU, some of the trajectories were generated in OpenMM with hydrogen reweighting, constrained length of all covalent bonds and the timestep of 5 fs,[14] and other trajectories were run in Amber with hydrogen reweighting, constrained length of covalent bonds with hydrogen atoms and the timestep of 4 fs.[15] Coordinates of all atoms were recorded every 0.2 ns. The following checks of the resulting trajectories were performed: stability of the volume, potential and total energy of the system; the distance between mirror images of the (glyco)protein created by periodic boundary conditions (in more than 80% of frames exceeded 19 Å, 16 Å, 25 Å and 14 Å for the glycosylated GluN1 LBD, non-glycosylated GluN1 LBD, glycosylated GluN2B LBD and non-glycosylated GluN2B LBD, respectively; in more than 99% of frames, exceeded 15, 11, 16 and 9.5 Å, respectively); RMSD of the protein backbones in neighboring frames in each MD trajectory (always stayed below 4.5 Å for the timestep of 0.2 ns, and typically equaled ~1.5 Å). Visual molecular dynamics (VMD) package[16] was used to visualize molecular structures.

**Interpretation of molecular dynamics trajectories.** Markov state models[17] were built to reconstruct the thermodynamic and kinetic properties of the NPT ensembles of glycosylated and non-glycosylated GluN1 LBD and GluN2B LBD proteins at equilibrium from finite-length MD trajectories, each of which did not completely sample the configuration space. For featurization, the distance $d$ between $C_\alpha$ atoms in residues 507 and 701 in GluN1 LBD was used to capture the opening/closing motion of the module. This choice of the residues followed that used in the experimental papers on smFRET investigation of GluN1 LBD opening/closing dynamics on the millisecond-to-second timescale.[18,19] In GluN2B LBD, the distance between $C_\alpha$ atoms in residues 503 and 701, which are homologous to residues 507 and 701 in GluN1 subunit, was used. The MSMbuilder 3.3 package[20] was used to construct microstate models with varying number of Markov states. Maximum Likelihood Estimator was used to generate a transition probability matrix $T_{ij}$, which maps out the probability of transitioning from state $i$ at time $t$ to state $j$ at time $t+\tau$, where $\tau$ is the lag time of the model. The optimal number of Markov states was chosen by cross-validation with the use of generalized matrix Rayleigh quotient (GMRQ) for the slowest implicit timescale as the score function (SI, section S1).[1] The Markov lag time was chosen to be 256 ns based on the analysis of the plots for the implied timescales versus lag times used to compute these implied timescales (SI, section S2). MSM-weighted probability distributions were obtained by binning the raw data within each MSM state and weighting it by the MSM equilibrium state population. To estimate the stability of our key conclusions relative to the model framework used to process the MD data, we also present the results for MSMs with a different number of clusters or a different lag time, as well as the timescale estimated from time-structure independent component analysis (tICA),[21] all with the same featurization $d$ (Table 1).

## References for the Supplementary Information